\documentclass[twocolumn,graphicx,superscriptaddress,showkeys,floatfix]{revtex4-1} 
\usepackage{amsfonts}
\usepackage{graphicx}
\usepackage{bm}
\usepackage{dcolumn}
\usepackage{amssymb}
\usepackage[colorlinks,linkcolor=blue,citecolor=blue]{hyperref}
\usepackage{epsfig}
\usepackage{multirow}
\usepackage{amsmath}
\usepackage{hhline}
\usepackage{mathptmx}
\usepackage{eucal}
\usepackage{color}
\usepackage{epstopdf}
\usepackage{natbib}
\usepackage{gensymb}
\pdfpageattr {/Group << /S /Transparency /I true /CS /DeviceRGB>>}

\begin{document}
\title{Pb/InAs nanowire Josephson junction with high critical current and magnetic flux focusing}

\author{J. Paajaste}
\email{jonna.paajaste@nano.cnr.it}
\affiliation{NEST, Istituto Nanoscienze-CNR and Scuola Normale Superiore, I-56127 Pisa, Italy}

\author{M. Amado}
\affiliation{NEST, Istituto Nanoscienze-CNR and Scuola Normale Superiore, I-56127 Pisa, Italy}

\author{S. Roddaro}
\affiliation{NEST, Istituto Nanoscienze-CNR and Scuola Normale Superiore, I-56127 Pisa, Italy}

\author{F. S. Bergeret}
\affiliation{Centro de Fisica de Materiales (CFM-MPC), Centro Mixto CSIC-UPV/EHU, E-20018 San Sebastian, Spain}
\affiliation{Donostia International Physics Center (DIPC), E-20018 San Sebastian, Spain}

\author{D. Ercolani}
\affiliation{NEST, Istituto Nanoscienze-CNR and Scuola Normale Superiore, I-56127 Pisa, Italy}

\author{L. Sorba}
\affiliation{NEST, Istituto Nanoscienze-CNR and Scuola Normale Superiore, I-56127 Pisa, Italy}

\author{F. Giazotto}
\email{f.giazotto@sns.it}
\affiliation{NEST, Istituto Nanoscienze-CNR and Scuola Normale Superiore, I-56127 Pisa, Italy}


\begin{abstract}
We have studied mesoscopic Josephson junctions formed by highly \textit{n}-doped InAs nanowires and superconducting Ti/Pb source and drain leads. The current-voltage properties of the system are investigated by varying temperature and external out-of-plane magnetic field. Superconductivity in the Pb electrodes persists up to $ \sim 7$ K and with magnetic field values up to 0.4 T. Josephson coupling at zero backgate voltage is observed up to 4.5 K and the critical current is measured to be as high as 615 nA. The supercurrent suppression as a function of the magnetic field reveals a diffraction pattern that is explained by a strong magnetic flux focusing provided by the superconducting electrodes forming the junction.

\end{abstract}


\maketitle

\section{Introduction}
\begin{figure*}[t!]
\includegraphics[width=0.9\textwidth,clip=]{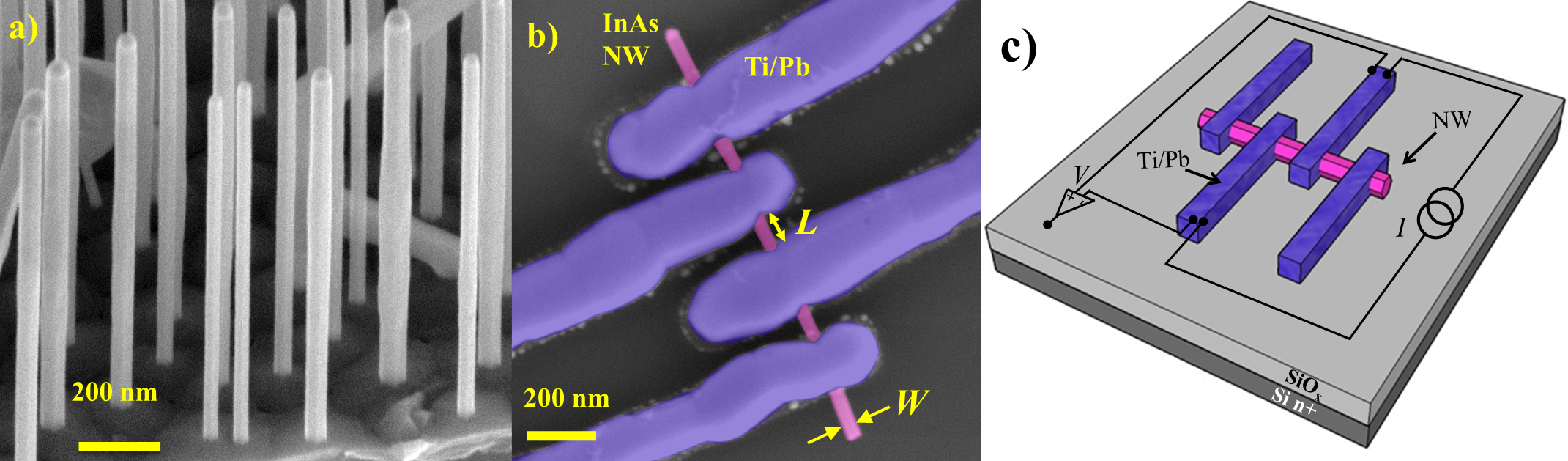}
\caption{a) Electron micrograph of a typical distribution of InAs nanowires after their growth. The image was taken at a 45\degree angle. b) False colored electron micrograph of our typical InAs nanowire-based Josephson junction of length \textit{L} and width \textit{W}. The nanowire appears in pale pink whereas the Ti/Pb leads in blue. c) A sketch of the final setup is displayed with the typical 4-wires configuration, where the Josephson junction is current biased and the voltage drop is measured with a room-temperature voltage preamplifier.}\label{NW-device-sketch}
\end{figure*}

The coupling of superconducting electrodes by a weak link can result in the formation of a Josephson junction, where a dissipationless supercurrent is induced due to the proximity effect, i.e., the penetration of superconducting correlations in the non-superconducting link resulting from Andreev reflection \cite{Gennes-1966, Andreev-1964, Giazotto-2010}. Josephson junctions and coherent quantum transport have been attracting interest in recent years due to the development of modern nanofabrication techniques, which enable the fabrication of nanoscale hybrid superconducting devices in a broad range of design and materials. Josephson junctions realized with semiconductor nanowires (NWs) \cite{DeFranceschi-2010,Doh-2005,Xiang-2006,Dam-2006, Frielinghaus-2010, Nishio-2011,Roddaro-2011} have shown a high potential in different types of nanoscale devices demonstrating, for example, tunable supercurrents \cite{Doh-2005,Xiang-2006}, the supercurrent reversal \cite{Dam-2006}, and the suppression of supercurrent by hot electron injection \cite{Roddaro-2011}. NWs have also been used for superconducting quantum interference devices (SQUIDs) \cite{Dam-2006} and tunable Cooper pair splitters \cite{Hofstetter-2009}. Benefits of using NWs include the tunable charge density by means of a gate voltage that in turn tunes the supercurrent \cite{Doh-2005}. Yet, small dimensions of NWs also make them a promising choice for studying fundamental phenomena such as the quantum interference \cite{Giazotto-2010} and Majorana fermions \cite{Mourik-2012, Xu-2012,Shtrikman-2012,Lee-2013}.

The vast majority of the studies of NW-based Josephson junctions rely on Al superconducting electrodes limiting the operation to maximum 1.2 K, which corresponds to the critical temperature $T_{C}$ of the Al-based leads. The use of superconductors with higher $T_{C}$ is required to extend the device operation capabilities to higher temperatures. This also results a higher critical magnetic field $ B_{C} $ which sets an upper limit to their operation at external magnetic fields. This will enable the use of these hybrid junctions not only in dilution refrigerators but also in He-liquid-based systems where the base temperature will be typically higher than the $T_{C}$ of Al. Proximity DC SQUIDs based on InAs NWs and vanadium (V) superconducting electrodes ($T_{C} \simeq$ 4.6 K) have been realized and show that V/InAs-NW/V Josephson junctions can operate at temperatures up to 2.5 K \cite{Spathis-2011,Giazotto-2011}. Other studies explore Josephson junctions with Nb contacts ($T_{C} \simeq$  9.3 K) showing a supercurrent up to 4 K with InAs NWs \cite{Gunel-2012} while by using NWs such as InN a supercurrent was observed up to temperatures of 3.5 K \cite{Frielinghaus-2010}. 

In this work, we report an exhaustive study of the response of InAs NW Josephson junctions with Pb ($T_{C} \simeq$ 7.2 K) superconducting leads as a function of the temperature and in the presence of an external perpendicular magnetic field. To our knowledge Pb has not been used previously in InAs or any NW Josephson junctions, though a dilute PbIn compound and Pb have proved to be a suitable candidates for Josephson junctions using graphene as a weak link \cite{Jeong-2011, Borzenets-2012}. The high critical temperature of Pb combined with the ease of evaporation makes it a promising and solid alternative to superconducting Nb as a contact electrode material.

As well as the choice of materials, the dimensions and geometry affect the performance of mesoscopic Josephson junctions. In principle, for junctions with the ratio $W/L$ (\textit{W} is the width and \textit{L} the length of the junction) smaller than unity, the critical supercurrent is expected to decrease monotonically as a function of the external perpendicular magnetic field \cite{Bergeret-2008}, instead of showing a Fraunhofer diffraction pattern typical of wide junctions \cite{Frielinghaus-2010,Gunel-2012, Abay-2012}. Although our system has dimensions of a narrow junction, with the width smaller than the coherence length \cite{Cuevas-2007}, we observe a magnetic diffraction pattern which can be explained in terms of a strong magnetic focusing provided by the superconducting electrodes forming the junction \cite{Gu-1979}.

\section{The sample fabrication}


InAs NWs used in this work have been grown by Au-assisted chemical beam epitaxy (CBE) \cite{Vitiello-2012, Ercolani-2009} on InAs (111)B substrates. Trimethylindium (TMIn) and tertiarybutylarsine (TBAs), which was cracked at 1000 \degree C, were employed as metal organic (MO) precursors for InAs growth and ditertiarybutyl selenide (DtBSe) as a selenium source for \textit{n}-type doping. Prior to NW growth a thin Au film was deposited on the InAs wafer and the wafer was then annealed at 520 \degree C under TBAs flow in order to remove the surface oxide from the InAs substrate and generate the Au nanoparticles by thermal dewetting. The InAs segment was grown for 90 min at a temperature of (420 $\pm$ 5) \degree C, with MO line pressures of 0.3 and 0.7 Torr for TMIn and TBAs, respectively. For \textit{n}-type doping, the DtBSe line pressure was set to 0.4 Torr to achieve high carrier density, which is estimated to be $n_{s} \sim 3 \times 10^{18} $ cm$^{-3}$ and electron mobility $\mu \sim 2000$ cm$^{2}$/Vs based on previous experiments done with similar NWs \cite{Viti-2012}.

Figure 1(a) shows a micrograph image of the as-grown NWs obtained with a field emission scanning electron microscope (SEM) operating at 5 kV, with the sample tilted 45\degree. The NW diameters are in the 50--80 nm range, and have a typical length of approximately 800 nm. The NWs show some lateral overgrowth with irregular faceting which is typical for highly Se-doped InAs NWs \cite{Viti-2012,Thelander-2010}. The presence of DtBSe during the growth both increases the zincblende fraction \cite{Thelander-2010} and reduces the adatom surface diffusion leading to lateral growth and new facet formation \cite{Viti-2012}.


Coarse Ti/Au pads and fine alignment markers were evaporated on an \textit{n}-Si/SiO$_{2}$ substrate prior to the mechanical transfer of the NWs. Two subsequent steps of SEM imaging and self-aligned electron beam lithography (EBL) were performed in order to define Josephson junctions with Ti/Pb leads and InAs NW as the weak link. Before the last metallization the sample was chemically etched with an ammonium polysulfide (NH$_{4}$)$_{2}$S$_{x}$ solution in order to remove the native oxide layer of the NWs, to reduce surface scattering processes, and to ensure good ohmic contacts \cite{Sourribes-2013}. The metal deposition of the Ti/Pb (10/80 nm) leads was performed with deposition rate of 4 \r{A}/s at room temperature by electron beam evaporation under ultra-high vacuum conditions. The highly volatile nature of Pb makes it easy to evaporate but it is also more difficult to control on the surface since it forms islands of Pb rather than wetting the surface. A 10-nm-thick Ti layer is used as a wetting layer whereas the Pb represents the superconducting lead of the Josephson junction. Figure 1(b) shows a typical final device with Ti/Pb leads with an interelectrode spacing of \textit{L}$\sim$100 nm. It can be seen how instead of sharp and well-defined Pb interfaces, which are typical for the high energy EBL, the Pb electrodes have a rounded shape. This is due to the fact that Pb tends to retract on top of the Ti wetting layer and also from the NW in a droplet-like fashion. Parts of Ti wetting layer can in fact be seen surrounding the electrodes in Fig. 1(b). The issue of retraction could be overcome only thanks to the carefully-selected deposition rate and layer thickness. The results obtained in this study clearly demonstrate that the deposition of Pb can be done with a normal room-temperature sample holder, and that Pb presents a suitable candidate for hybrid superconducting devices with nanowires as a weak link.

\section{Experimental results and discussion}


The hybrid NW-based Josephson junctions were characterized in a filtered dilution refrigerator down to 10 mK. Even if four superconducting electrodes are present in the device, only a single neighbouring pair was used during the transport measurements. The structure was biased by a current $ I $, whereas the voltage drop across the NW has been registered via a room-temperature-differential-preamplifier [see Fig. 1(c)]. A full characterization of the electrical response of the Josephson junction as a function of the temperature and out-of-plane external magnetic field (\textit{B}) was performed. 

Figure 2(a) shows the typical current voltage (IV) characteristics of our Ti/Pb/InAs NW Josephson junction measured at different temperatures. A sizeable supercurrent of 615 nA has been observed at the base temperature (10 mK) yielding a supercurrent density of $\sim31.3$ kA/cm$^{2}$. Without any external applied magnetic field, the supercurrent persists up to 4.5 K. Such high value of operational temperature can be attributed to the use of Pb as superconducting source and drain electrodes as well as to the limited length of the junction. Compared to previously reported InAs and InSb NW Josephson junctions with Al and Nb contacts, the measured maximum supercurrent of our device is about twice as large \cite{Roddaro-2011,Gunel-2012, Nilsson-2012}. Only in recent works with very short ($ \sim 30$ nm) Al/InAs-NW Josephson junctions a higher supercurrent of $ \sim 800$ nA has been measured with highly-transparent contacts \cite{Abay-2012, Abay-2014} and in InN-NW/Nb Josephson junctions even an order of magnitude larger supercurrent has been observed \cite{Frielinghaus-2010}. At temperatures below 1.5 K a remarkable hysteretic behaviour between the switching $I_{Cs}$ and retrapping $I_{Cr}$ currents is evident as seen in Fig. 2(a). Such hysteresis stems from quasiparticle heating when the junction switches from the normal to the superconducting state \cite{Courtois-2008}, and has been previously reported in a number of hybrid Josephson junctions with various geometries of the weak-link \cite{Fornieri-2013,Amado-2013,Gunel-2012}.


\begin{figure}[t!]
\centerline{\includegraphics[width=0.5\textwidth,clip=]{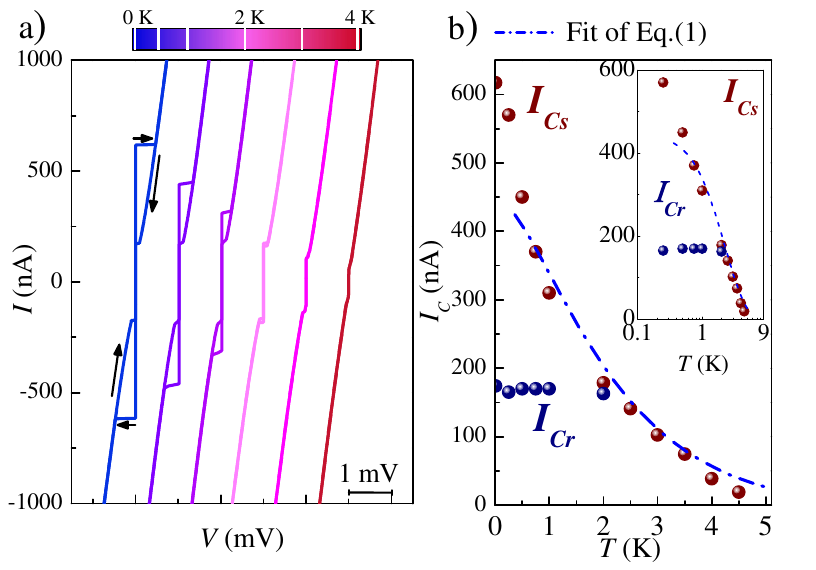}}
\caption{a) Back and forth current vs voltage characteristics (IVs) for Pb/InAs NW/Pb junction with 100 nm electrode spacing measured at different temperatures indicated in the color bar. The curves are horizontally offset by 1 mV for clarity. At T $ < $ 1.5K quasiparticle heating dominates the transition from superconducting to normal state resulting in a strong hysteretic behavior which depends on the direction of the bias, as indicated by arrows. b) Switching $ I_{Cs} $ and retrapping currents $ I_{Cr} $ extracted from the IVs as a function of the temperature, with the sphere radius indicating the estimated error. As it can be seen, $ I_{Cs} $ decays exponentially at temperatures higher than 250 mK. The data is fitted with the proposed model described by Eq.(1) (dash-dot line) with parameters \textit{L}=350 nm, \textit{D}=0.02 $ m^{2}/s $, $R_b$=82 $ \Omega $, and  $R_{NW}$=411 $ \Omega $. The inset shows the evolution of the $ I_{Cs} $ and $ I_{Cr} $ in normal-log scale.}
\label{VI}
\end{figure}

The expected exponential decay of $I_{Cs}$ at high temperatures and the saturation at $T\rightarrow 0$  is presented in Fig. 2(b) \cite{Dubos-2001}. To fit the experimental data for $I_c$ at high temperatures we use the expression for a SNS junction (where N is a normal metal and S is a superconductor) obtained  by solving the linearized Usadel equation \cite{Kubrianov-1988}:
 \begin{equation}
 I_C=\frac{\pi k_{B} }{eR_{NW}} T\sum_{\omega}\frac{(\kappa_\omega L)\Delta^2  }{ (\omega^2+\Delta^2)\left[\alpha\sinh(\kappa_\omega L)+\beta \cosh(\kappa_\omega L)\right]}
 \end{equation}
where the sum is over the Matsubara frequencies $\omega=\pi k_{B}T(2n+1)$, $n=0,\pm1,\pm2, ...$, $ k_{B} $ is the Boltzmann constant, $\kappa_\omega=\sqrt{2|\omega|/(\hbar D)}$, \textit{D} is the wire diffusion coefficient, $ \hbar $ is the reduced Planck constant, $R_{NW}$ is the resistance of the wire of length \textit{L}, $ \alpha=1+r^2(\kappa_\omega L)^2$, $ \beta=2r(\kappa_\omega L) $, and $r=R_b/R_{NW}$ with $R_b$ being the resistance of the S/N contact. A fit of Eq.(1) is presented in Fig. 2(b) and it corresponds to values \textit{L}=350 nm, \textit{D}=0.02  m$^{2}$/s , $R_b$=82  $ \Omega $, and  $R_{NW}$=411  $ \Omega $. The fit indicates that the effective length of the junction ($ \sim 350 $ nm) is much larger than the interelectrode spacing. The geometry of the device supports this observation, since the electrodes cover a considerable section of the NW.

The numerical differential resistance \textit{dV/dI} was calculated from the IVs measured as a function of perpendicular-to-the-plane magnetic field at several temperatures ranging from 15 mK to 6 K. From the \textit{dV/dI} characteristics shown in Fig. 3(a) we can determine the superconducting energy gap of the electrodes to be $\Delta_{0} \sim$ 1.37 meV. The value is based on features corresponding to $n=1$ and $n=2$ (labelled in Fig. 3(a)) of multiple Andreev reflections, $V_{n}=2\Delta_{0}/ne $, where $ e $ is the elementary charge and $ n $ an integer. The junction total normal-state resistance $R_{N}$ is also estimated from the \textit{dV/dI} characteristics yielding a value \textit{R$_{N}$} $\sim$ 575 $ \Omega$ at 15 mK and resulting an $I_{C}R_{N}$ product as high as $\sim350$ $\mu$eV. Considering the sizeable supercurrent in our junction and the low contact resistance from the simulation, we estimate the quality of the interfaces to be very good.

\begin{figure}[t!]
\centerline{\includegraphics[width=0.5\textwidth,clip=]{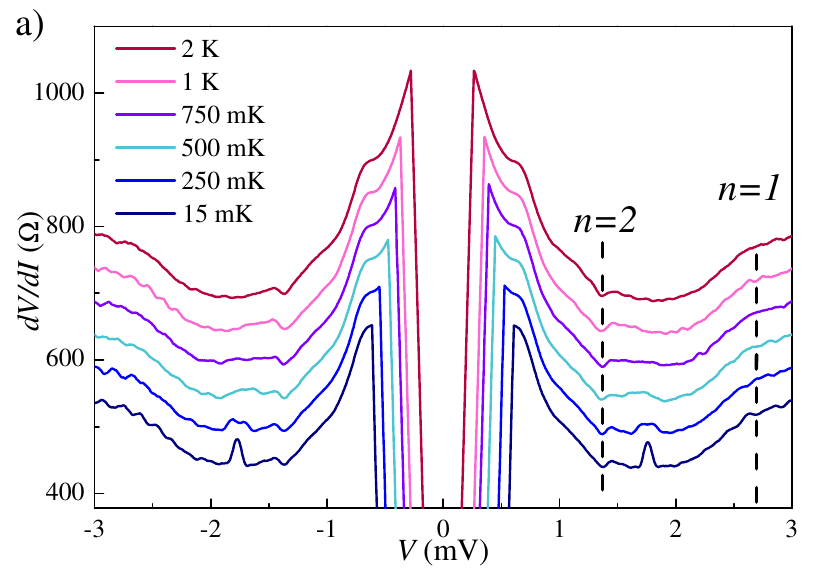}}
\centerline{\includegraphics[width=0.5\textwidth,clip=]{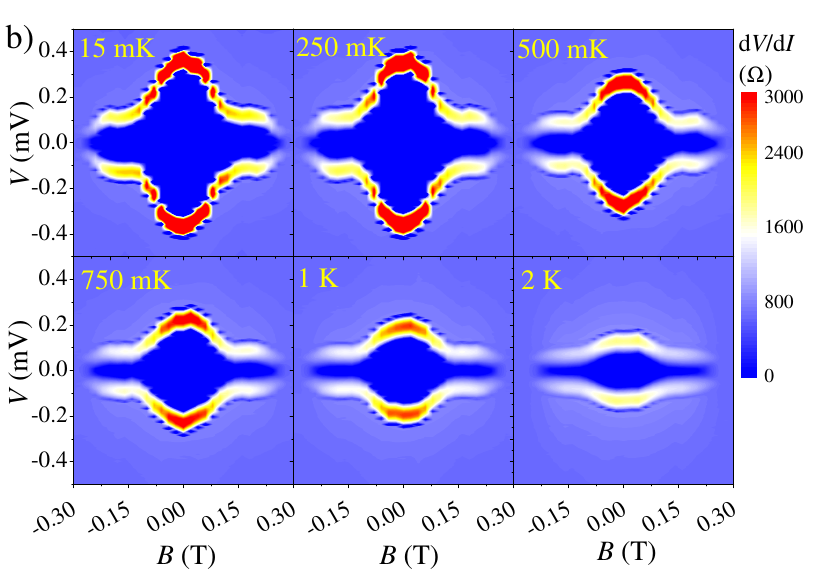}}
\caption{a) Differential resistance \textit{dV/dI} calculated numerically from IVs measured at zero magnetic field and different temperatures. Features related to multiple Andreev reflection and the superconducting gap of Pb are indicated with dashed lines. The curves are vertically shifted by 50 $\Omega$ for clarity. b) Contour plots of the differential resistance \textit{dV/dI} as a function of voltage and perpendicular magnetic field at different temperatures.}\label{dVdI}
\end{figure}

Based on field effect-measurements performed on almost identically grown NWs \cite{Viti-2012} we have estimated the electron mean free path to be $ \l_{e}=\nu_{F}m^{*}\mu/e \approx$ 60 nm, where the Fermi velocity for  the NW is $ \nu_{F}=\hbar \sqrt[3]{3 \pi^{2}n_{S}}/m^{*} $, the effective electron mass for InAs NW is $ m^{*}=0.023m_{e}$, and $ m_{e} $ is the electron mass.  The diffusion coefficient of the NW calculated based on a previous room-temperature experiment is $ D=\l_{e}\nu_{F}/3 \approx$ 0.046 m$ ^{2} $/s \cite{Abay-2012}, which is twice the value extracted from the earlier fit. The difference might be due to the fact that the values of mobility and carrier concentration were obtained for another batch of nanowires and were extracted from measurement of transistor devices with long junctions \cite{Viti-2012}. In our device the portion of the NW which has been chemically etched to allow the subsequent contact with the superconducting leads is comparable to the junction length. Therefore, we expect to have a substantial portion of the Josephson junction characterized by a larger disorder and enhanced scattering which would explain the smaller value of $D$ obtained from Eq. (1). We estimate the junction Thouless energy to be $ E_{th}=\hbar D/L^{2} \approx$ 3.0 meV with the junction length \textit{L} = 100 nm. The above obtained values seem to indicate that our device is a diffusive junction with an  intermediate length, since $\xi_{0}>L> \l_{e}$ and $ \Delta_{0} < E_{Th}$, where the coherence length is $ \xi_{0} = \sqrt{\hbar D/ \Delta_{0}}\approx$ 150 nm. However if we consider the values obtained from the fit of Eq.(1), we get $ \Delta_{0} > E_{th}\approx$1.0 meV and $ \xi_{0}\approx$100 nm which in turn point to the \textit{long} junction regime if we consider that the effective junction length is $\sim$350 nm, i.e., $ \simeq 3.5 \xi_{0} \ $. This conclusion is supported as well by the observed strong exponential damping of the critical current $ I_C $ as a function of temperature displayed in Fig. 2(b).

The full evolution of the \textit{dV/dI} as a function of the voltage and the magnetic field for temperatures from 15 mK to 2 K is presented in Fig. 3(b). In the contour plots a symmetric decay of the supercurrent (indicated by the blue regions) as a function of the magnetic field is clearly visible even for the higher temperatures at which the supercurrent starts to decrease. The supercurrent suppression itself has an  interesting shape and several features that attract a more detailed inspection. 

The evolution of the maximum critical current $I_{C}$ as a function of the perpendicular-to-the-plane external magnetic field is displayed in Fig. 4(a) for different temperatures. The magnetic field is varied from -0.3 T to 0.3 T while temperature is spanned from 10 mK to 4 K. As it can be observed, for temperatures as high as 2 K the supercurrent survives up to $ B\sim $0.28 T, a value much higher than the critical magnetic field of pure Pb bulk superconductor ($\sim0.08$ T) \cite{Martienssen-2005}. The critical magnetic field, in fact, can be enhanced drastically for thin evaporated layers due to the formation of grain boundaries in polycrystalline materials \cite{Meservey-1971}. Figure 4(b) shows the hysteretic nature of the maximum critical current at low temperature (10 mK) with respect to the magnetic field sweep direction. The observed switching behavior at 10 mK is attributed to possible trapping of some fixed number of fluxons (Abrikosov vortices) in the leads, which results a telegraphic noise as the junction switches back and forth between two states. \cite{Han-1989} The hysteretic behavior we attribute to pinning of these states, as predicted to occur in the presence of impurities \cite{Abrikosov-1957}.

\begin{figure}[t!]
\centerline{\includegraphics[width=0.5\textwidth,clip=]{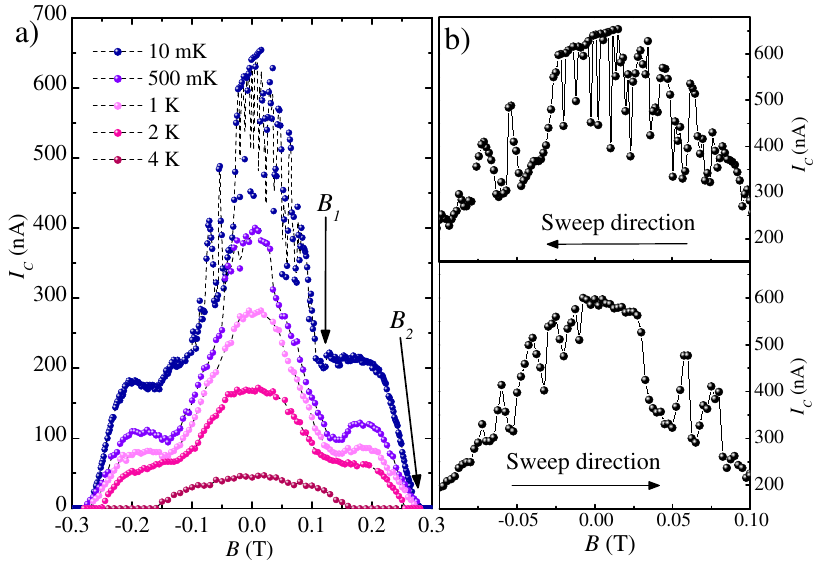}}
\caption{a) The critical current $ I_{C} $ as a function of perpendicular magnetic field \textit{B} for different temperatures from 10 mK to 4 K. Each point represents one IV measurement where the magnetic field and temperature have been kept constant. The beginning and the end points of the shoulder structure $ B_{1} $=0.13 T and $ B_{2} $=0.28 T are indicated by arrows. b) A blow up of the 10 mK $ I_{C} $ as a function of magnetic field shows the hysteretic behavior depending of the direction of the magnetic field sweep.}\label{Ic-B}
\end{figure}

A remarkable feature of the magnetic field suppression of the supercurrent is the presence of a shoulder starting at a magnetic field value $B_{1}$=0.13 T and ending at $B_{2}$=0.28 T, as indicated in Fig. 4(a). In the following, we will show that the observed behavior is consistent with a Fraunhofer-like diffraction pattern occurring in the Josephson junction. Typically in the case of a narrow junction the formation of vortices is not favorable and the field acts as a pair-breaking mechanism therefore suppressing the critical current monotonously \cite{Cuevas-2007,Bergeret-2008}. In our device the junction width, i.e., the diameter of the NW is 50 nm which corresponds to $\sim0.3\xi_{0}$ and in terms of magnetic length ($\xi_{B}=\sqrt{\Phi_{0}/B_{1}}=120$ nm), it is  $\sim0.4\xi_{B}$. Such values point towards a narrow junction \cite{Cuevas-2007,Bergeret-2008} for which we would expect a monotonic decay of the supercurrent. By contrast, the shape shown in Fig. 4(a) resembles the well-kown Fraunhofer diffraction pattern typical of wide junctions.
%
%

The calculated magnetic fluxes $\Phi=BA$ with our junction area $ A=LW $, interelectrode spacing \textit{L} = 100 nm, NW width \textit{W} = 50 nm, at points  $B_{1}$ and $B_{2}$ lie at $\Phi_{1}=0.7 \times 10^{-15} $ Wb$= 0.3\Phi_{0}$ and $\Phi_{2}=1.4 \times 10^{-15} $Wb$= 0.7\Phi_{0}$, where  $\Phi_{0}=h/2e = 2.07 \times 10^{-15} $ Wb is the flux quantum. Clearly the fluxes $\Phi_{1}$ and $\Phi_{2}$ do not fit the Fraunhofer diffraction pattern minima which are expected to occur at $\Phi_{0}$ and $2\Phi_{0}$. In order to explain the features of measured $I_C(B)$ curves we have to take into account the existence of flux focusing in the junction due to the fact that superconducting leads repel the external magnetic field. In order to accommodate the magnetic flux focusing into the diffraction pattern, we have to include in the Fraunhofer pattern 

\begin{equation}
\dfrac{I_{C}}{I_{C0} }=\dfrac{sin\pi C \Phi/\Phi_{0}}{\pi C \Phi/\Phi_{0}},
\end{equation}
a magnetic focusing factor $C=B_{Eff}/B_{Ext}$, where $ B_{Eff} $ is the effective magnetic field at the center of the junction and $ B_{Ext} $ is the external applied magnetic field perpendicular to the sample surface. The parameter \textit{C} represents the magnetic focusing factor and, when $ > 1$, it scales the Fraunhofer pattern to smaller $\Phi/\Phi_{0}$ values.

\begin{figure}[t!]
\centerline{\includegraphics[width=0.5\textwidth,clip=]{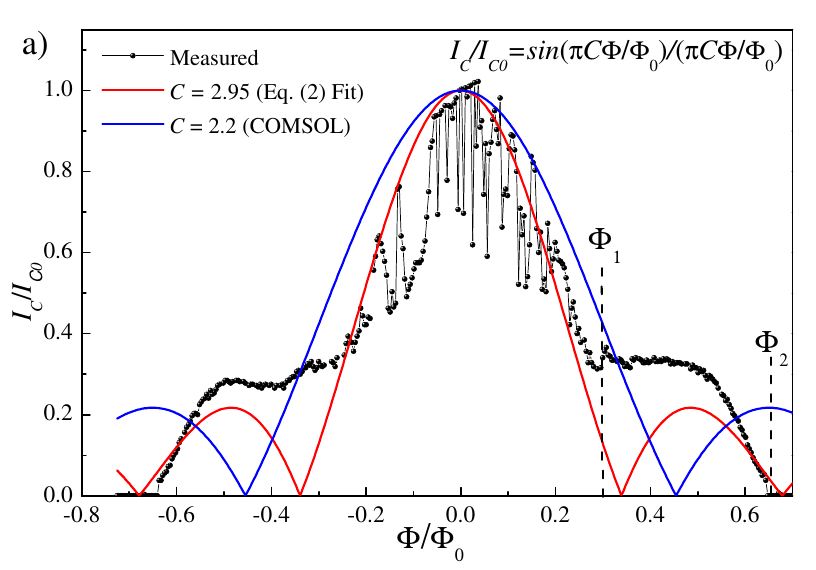}}
\centerline{\includegraphics[width=0.5\textwidth,clip=]{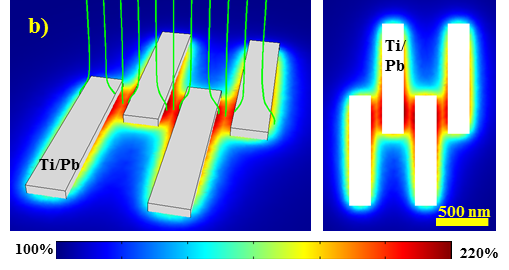}}
\caption{a) The measured normalized critical current versus normalized magnetic flux. Solid lines are Fraunhofer patterns according to Eq. (2) corrected with a magnetic flux focusing parameter \textit{C} obtained from a COMSOL simulation (blue) and a fit of Eq. (2) to the data (red). b) A COMSOL simulation results showing the strength of perpendicular magnetic field in our system depending on the position.}\label{Fraunhofer}
\end{figure}

To estimate the magnitude of magnetic flux focusing effect, i.e., the value of \textit{C}, we have used two approaches, and  Fraunhofer patterns corresponding to both methods are shown along with the measured data in Fig. 5(a). First, using the COMSOL program with Matlab livelink, we have simulated our device as an array of rectangular electrodes [Fig. 5(b)] in a cubic space (one side 4 $\mu$m) much larger than the Josephson junction in order to minimize spurious effects due to the finite size of the simulation. On the electrode surfaces we used magnetic insulation i.e. full Meissner effect as a boundary condition. The cube surfaces in lateral directions were also set as magnetic insulating. The surfaces in vertical direction (the direction of out-of-plane field) had a boundary condition of constant magnetic potential to create a static and uniform magnetic field. The simulation reveals an effective magnetic field of $B_{Eff}=2.2 \times B_{Ext}$ which gives us a magnetic focusing factor $C=B_{Eff}/B_{Ext}=2.2$. The second method used to estimate magnetic flux focusing is the fit of equation (2) to obtain the optimal value of $ C $. This results in $C = 2.95$, a value which is comparable to the one obtained from COMSOL simulation, so clearly both methods support the focusing assumption. For this value of $ C $, we get $\xi_{B}=\sqrt{\Phi_{0}/B_{Eff}}=\sqrt{\Phi_{0}/(CB_{1})} =70$  nm, which is comparable to \textit{W} and smaller than \textit{L}, thus supporting the existence of magnetic diffraction.




Another interesting feature of our experimental data is the deviation of the minimum at $ \vert B_{1} \vert $ from the expected zero of the theoretical Fraunhofer diffraction pattern. This deviation can be explained if we consider also the intrinsic properties of our junction. It is  known that InAs NWs have surface states that accumulate surface charge \cite{Smit-1989} thereby causing a nonuniform current density distribution. Furthermore, in the case of a heavily-doped NW, the mixed wurzite and zincblende crystalline structure is likely to affect the current density distribution. Ultimately both effects lead to a nonunifrom supercurrent density distribution, and can cause a disturbance in the complete destructive interference \cite{Barone}.

\section{Summary}

In summary, we have successfully fabricated mesoscopic Pb/InAs-NW/Pb Josephson junctions with $\sim100$ nm interelectrode spacing by a room-temperature Pb evaporation. Devices show excellent properties in terms of a high critical current exceeding 600 nA and an observable Josephson current at temperatures up to 4.5 K and magnetic fields up to 0.3 T. Superconductivity is observed in the electrodes up to $ \sim $7 K. Unlike the expected behaviour in narrow junctions where the width is smaller than the magnetic length, we observe a marked diffraction pattern of the critical current as a function of the magnetic field which can be explained by a strong magnetic flux focusing provided by the Pb electrodes.\\

\section{ACKNOWLEDGEMENTS}

Partial financial support from the Marie Curie Initial Training Action (ITN) Q-NET 264034 and the
Tuscany region through the project "TERASQUID" is acknowledged. The work of F.G. has been partially funded by the European Research Council under the European Union's Seventh Framework Programme (FP7/2007-2013)/ERC grant agreement No. 615187-COMANCHE. The work of F.S.B. was supported by the Spanish Ministry of Economy and Competitiveness under Projects No. FIS2011-28851-C02-02.


\end{document}